\documentclass[prb,aps,twocolumn,superscriptaddress]{revtex4-1}

\usepackage{bm}
\usepackage[colorlinks=true,linkcolor=blue,citecolor=blue]{hyperref}
\usepackage{times}
\usepackage{amsmath}
\usepackage{amssymb}
\usepackage{amsthm}
\usepackage{amsfonts}
\usepackage{enumerate}
\usepackage{latexsym}
\usepackage{ifpdf}
\newcommand{\beq}{\begin{equation}}
\newcommand{\eeq}{\end{equation}}
\usepackage{graphicx}
\usepackage{makeidx}
\usepackage{color}

\begin{document}

\title{Epitaxial strain modulated electronic properties of interface controlled nickelate superlattice}
\author {S. Middey}
\email{smiddey@iisc.ac.in  }
\affiliation  {Department of Physics, Indian Institute of Science, Bangalore 560012, India}
\author{D. Meyers }
\affiliation  {Department of Condensed Matter Physics and Materials Science, Brookhaven National Laboratory, Upton, New York 11973, USA}
\author {Shashank Kumar Ojha}
\affiliation  {Department of Physics, Indian Institute of Science, Bangalore 560012, India}
\author{M. Kareev}
\affiliation  {Department of Physics and Astronomy, Rutgers University, Piscataway, New Jersey 08854, USA}
\author{X. Liu}
\affiliation  {Department of Physics and Astronomy, Rutgers University, Piscataway, New Jersey 08854, USA}
\author{Y. Cao}
\affiliation  {Department of Physics and Astronomy, Rutgers University, Piscataway, New Jersey 08854, USA}
\author{J. W. Freeland}
\affiliation {Advanced Photon Source, Argonne National Laboratory, Argonne, Illinois 60439, USA}
\author{ J. Chakhalian}
\affiliation  {Department of Physics and Astronomy, Rutgers University, Piscataway, New Jersey 08854, USA}

\begin{abstract}

Perovskite nickelate heterostructure consisting of single unit cell of EuNiO$_3$ and LaNiO$_3$ have been grown on a set of single crystalline substrates by pulsed laser interval deposition to investigate the effect of epitaxial strain on electronic and magnetic properties at the extreme interface limit. Despite the variation of substrate in-plane lattice constants and lattice symmetry, the structural response to heterostructuring is primarily controlled by the presence of EuNiO$_3$ layer. In sharp contrast to bulk LaNiO$_3$ or EuNiO$_3$, the superlattices grown under tensile strains exhibit metal to insulator transition (MIT) below room temperature. The onset of magnetic and electronic transitions associated with the MIT can be further separated by application of large tensile strain. Furthermore, these transitions can be entirely suppressed by very small compressive strain. X-ray resonant absorption spectroscopy measurements reveal that such strain-controlled MIT is directly linked to strain induced self-doping effect without any chemical doping. 

  \end{abstract}

\maketitle

\section{Introduction}
 The sudden change in the electrical conductivity across the metal insulator transition (MIT) of complex oxides remains a topic of long-standing interest in condensed matter physics and materials science~\cite{mit_rmp}. Apart from the fundamental physics aspect of understanding the origin of MIT, a lot of attempts are being made towards the realization of next generation functional devices utilizing MIT~\cite{ramanathanreview,tokuranature,parkinscience}. Practical realization of such devices depends strongly on the ability to maintain sharp metal-insulator transition as the size reduction of the materials towards the nanometer thick device scale and epitaxial strain can significantly modify MIT~\cite{schlom,hwang,jakrmp}.
 
 As a prototypical example having MIT, massive efforts have been made over the last 5 years about the manipulation of the MIT of rare-earth perovskite nickelate ($RE$NiO$_3$) using external perturbation such as light, strain, electric and magnetic fields etc. (see Ref. ~\onlinecite{ownreview,ramanathan1,ramanathan2}  and references therein). Epitaxial strain i.e. mismatch of lattice constants between the single crystalline substrate and $RE$NiO$_3$, has been found to be very successful in manipulation of these transitions~\cite{jian_nc,triscone111,nno_badmetal,eno_prb,sno_prb,sno_aplmaterials,keimer_raman,  snojap,lnothoery,lnoarpes}. For example, the first-order metal to insulator transition (MIT) can be suppressed entirely by compressive strain. Though the MIT is accompanied by spin and charge ordering transitions and structural symmetry lowering in bulk NdNiO$_3$~\cite{nno_co,scagnoliprb,lorenzoprb,noorbitalordering,keimer_scattering},  the MIT and magnetic transition can be separated by tensile strain, leading to a paramagnetic insulating phase~\cite{jian_nc}. Surprisingly, charge ordering and symmetry lowering transitions are absent in ultra-thin NdNiO$_3$  films (thickness ~ 6 nm), grown under tensile strain~\cite{nno_upton,derek_nno}. Nickelates being a prototypical strongly correlated system, exhibit highly nontrivial transport properties in the metallic phase. One such frequently discussed phenomenon is the non-Fermi liquid (NFL) behavior of the metallic phase, and epitaxial strain is able to control scaling behavior (power exponents) of the NFL phase~\cite{jian_nc,nno_badmetal}. In addition, $RE$NiO$_3$  members have been combined with dielectric materials such as LaAlO$_3$, DyScO$_3$  etc. to study the effect of quantum confinement, and the responses of the orbital and spin degrees of freedom to heterostructuring and epitaxial strain~\cite{lno_keimer_nm,lno_jian_prb,keimer_prl13,keimer_xld13,lno_epl,lno_keimer_science,nno_prmaterials,keimer_xld15}. However, study of ultra-thin superlattices consisting of dissimilar nickelate layers is very limited~\cite{own_enolno}  and the response of electronic and magnetic structure to the underlying epitaxial strain is still largely unknown.

The choice of $RE$ ions determine the structural symmetry of  bulk $RE$NiO$_3$ and a very strong connection between the temperature scale of electronic, magnetic transitions and $<$Ni-O-Ni has been observed in bulk $RE$NiO$_3$ series~\cite{rno_phase1,rno_phase}. For example, bulk LaNiO$_3$ (LNO) with rhombohedral symmetry has the smallest distortion ($<$Ni-O-Ni $\sim$ 165.2$^\circ$) in the $RE$NiO$_3$ series and  remains metallic and paramagnetic without any structural transition. On the other hand, bulk EuNiO$_3$ (ENO) is  strongly distorted ($<$Ni-O-Ni $\sim$ 147.9$^\circ$) and   undergoes a first order MIT around 460 K with a charge ordering transition and structural transition and well separated magnetic transition (paramagnetic to $E'$-antiferromagnetic) at $\sim$200 K~\cite{bulk_eudopedlno}.   Since each $RE$NiO$_3$ member has a rather strong  propensity  for maintaining  bulk-like symmetry  even in thin film geometry~\cite{icheng_prb}, a strong structural competition can be anticipated in the ultra-thin limit for  the superlatices consisting of dissimilar nickelates layers, and  can in turn result in new electronic and magnetic phenomena.

Towards this  goal, we have synthesized and investigated the effect of epitaxial strain by growing 1 uc EuNiO$_3$/ 1 uc LaNiO$_3$ superlattices (1ENO/1LNO SL, uc= unit cell in pseudocubic setting, see Fig. 1(a)) on a variety of substrates. To elucidate the  microscopic effect of epitaxial strain on the structural, electronic, and magnetic properties of these superlattices,   X-ray diffraction (XRD), dc transport, Hall effect, resonant soft X-ray absorption spectroscopy (XAS) and X-ray linear dichroism (XLD) measurements have been performed.   Surprisingly, we have found that  in-spite of the strong  variation of substrate strain and symmetry,  the structural  response of the SLs in this  ultimate interface  limit is primarily governed by the ENO layer. The heterostructure grown under tensile strain undergoes  a MIT and a magnetic transition below room temperature, emphasizing   entire modulation of the electronic properties, sharply contrasted  to  the bulk ENO and LNO.  Moreover, by the judicious application of epitaxial strain these transitions can be made to occur simultaneously or separated with temperature or even entirely suppressed. Oxygen $K$ edge XAS measurements revealed that such a drastic change in the electronic behaviour is related to a strain induced self-doping effect~\cite{selfdoping,selfdopinglno}. Such manipulation of the electronic and magnetic transitions by the application of epitaxial strain highlights the remarkable power of heteroepitaxy in determining physical properties of perovskite nickelates. 

\section{Experimental details}
 [1EuNiO$_3$/1LaNiO$_3$]$\times$10 superlattices  [(1ENO/1LNO) SL], oriented along the pseudo cubic [0 0 1] direction were grown on a variety of single crystal substrates by pulsed laser interval deposition~\cite{misha_jap,nnoprl}  from polycrystalline stoichiometric EuNiO$_3$ and LaNiO$_3$ targets. The substrates used in this work: DyScO$_3$ (DSO), NdGaO$_3$ (NGO), LaAlO$_3$ (LAO), and YAlO$_3$ (YAO)  have been selected to avoid  polar discontinuity at the film/substrate interface~\cite{scirep}. The symmetry of the substrates and the corresponding expected strain values for ENO and LNO  are listed in Table-I. Growth of all samples were monitored by in-situ high pressure RHEED (reflection high energy electron diffraction). All films were grown at 620$^\circ$ C and 150 mTorr of oxygen pressure and were post annealed at growth temperature under 650 Torr pressure of pure oxygen.  XRD measurements   were carried out around the (0 0 2)  reflection of the substrate (pseudocubic notation) with a Panalytical XPert Pro materials research diffractometer (MRD). X-ray absorption spectra (XAS) of  Ni $L_{2,3}$ edges  were recorded   at the 4-ID-C beam line of  Advanced Photon Source (APS). $dc$ transport measurements, using a four probe  Van Der Pauw geometry,  were performed in  Physical Property Measurement System (PPMS Quantum Design). $I$-$V$ measurements confirmed ohmic behavior of all electrical contacts.

\begin{table}
\caption {Symmetry and in-plane pseudo-cubic lattice constants  ($a_{sub}$) of the  substrates  and the corresponding strain ($\epsilon$) for orthorhombic EuNiO$_3$ (3.806 \AA),  and rhombohedral LaNiO$_3$ (3.855 \AA).  The lattice constants for the bulk ENO and LNO have been obtained from Ref.~\onlinecite{bulk_eudopedlno}.}
\begin{tabular}{lcccc}
\hline
\hline
 Substrate     & Symmetry      &  $a_{sub}$  ({\AA}) & $\epsilon$ &  $\epsilon$  \\
                      &                      &                                &  for ENO & for LNO  \\ \hline
 YAlO$_3$ & Orthorhombic & 3.692 & -3.0\% & -4.2\%          \\
 LaAlO$_3$ & Rhombohedral & 3.794 & -0.3\% & -1.6\%   \\
 NdGaO$_3$ & Orthorhombic & 3.858 & +1.4\% & +0.1\%   \\
 DyScO$_3$& Orthorhombic & 3.955 &   +3.9\% & +2.6\% \\  \hline
\end{tabular}
\end{table}

\section{Results and discussion}

{\it Epitaxial growth:} Since the optimal growth conditions for ENO and LNO thin films are different~\cite{eno_growth,own_apl}, it is crucial to find out the mutually compatible growth conditions for the layer by layer epitaxial stabilization of both ENO and LNO layers to form high quality 1ENO/1LNO SL.   The sequence of layer by layer growth has been shown in Fig. 1(b). The variation of the intensity of the specular spot in the RHEED  pattern, recorded during the growth of 1ENO/1LNO superlattice on a NGO substrate,  has been plotted in Fig. 1(c). The full recovery of intensity after the deposition of each unit cell confirms the desired layer by layer stabilization of both EuNiO$_3$ and LaNiO$_3$. The RHEED image (Fig. 1(d)), taken after cooling the sample to room temperature, shows the  streak patterns of specular (0, 0) and off-specular  (0, $\pm$1) Bragg reflections, implying atomically smooth surface morphology. The presence of half order reflections (marked by  arrows) due to the in-plane doubling of the unit cell ~\cite{eno_growth}  (also observed  for superlattices grown on other substrates), confirm that superlattices have either  orthorhombic or monoclinic symmetry at room temperature.

\begin{figure}
\vspace{-0pt}
\includegraphics[width=0.49\textwidth] {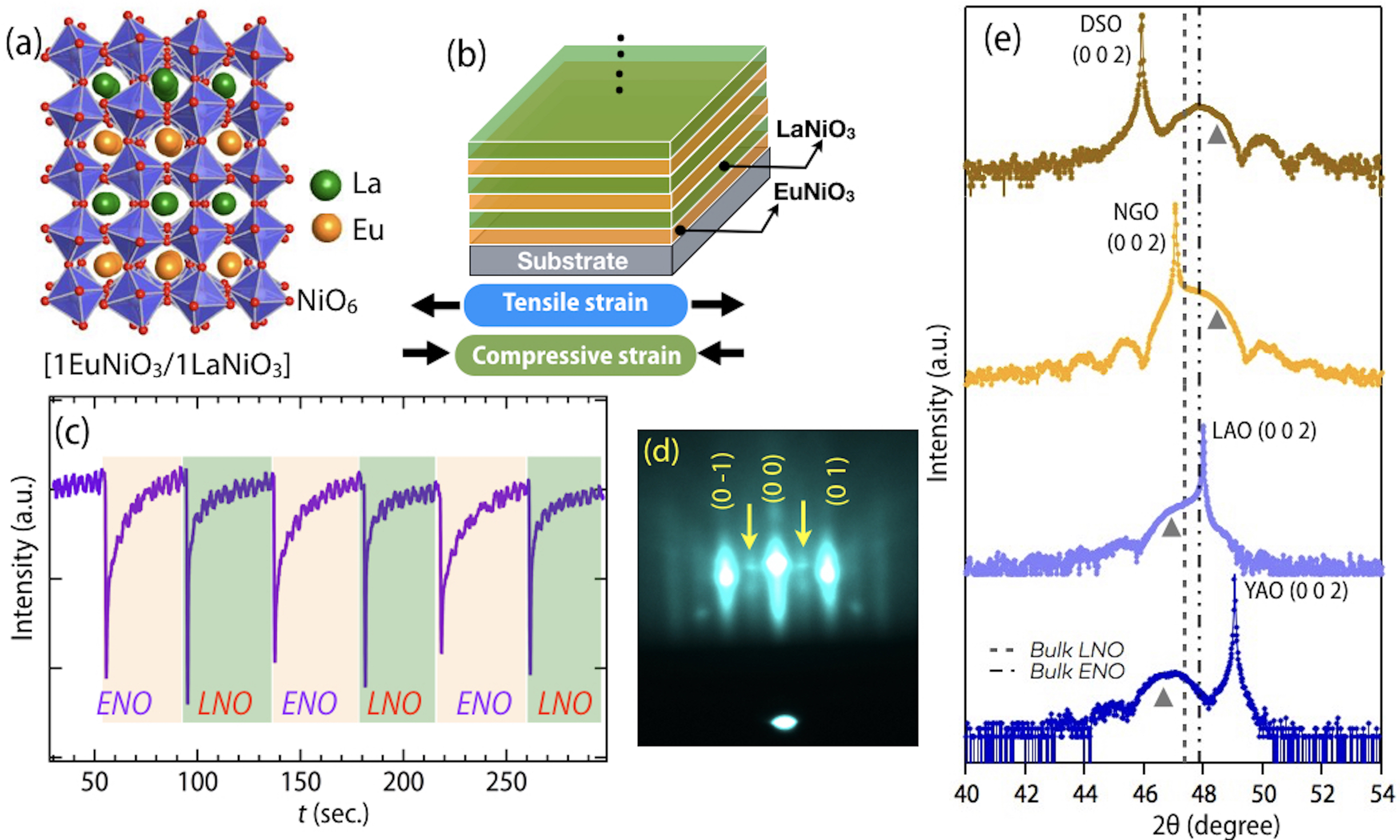}
\caption{\label{} (Color online) (a) 1EuNiO$_3$/1LaNiO$_3$ superlattice along pseudocubic [0 0 1]. (b) Schematic of sequence for layer-by-layer deposition used for this study.  (c) Intensity variation of RHEED specular spot during the deposition on NdGaO$_3$ substrate. (d) Final RHEED pattern  obtained along the [1 -1 0]  direction of NGO after cooling the film to room temperature. (e)  XRD patterns of [1ENO/1LNO]x10 superlattices on various substrates.  The  curves have been shifted along y-axis for visual clarity. The  vertical lines represent  the expected (0 0 2)$_{pc}$ peak position for   bulk EuNiO$_3$ and LaNiO$_3$.}
\end{figure}

Following Poisson argument about elasticity, it is generally anticipated that the single crystalline thin film should undergo out-of-plane compression (expansion) to accommodate in-plane tensile (compressive) strain.  Experimentally, however, the effects of epitaxial strain on nickelate thin films and heterostructures are  more complex  and markedly depart from the expected tetragonal distortion~\cite{eno_prb,jak_prl,icheng_prb,keimer_xld13,keimer_xld15}.  To investigate the effect of epitaxial strain on  our 1ENO/1LNO SLs,  2$\theta$-$\omega$ scans  have been recorded  using Cu $K_\alpha$ radiation (Fig. 1(e)). Each of the diffraction patterns consist of a sharp substrate peak together with a film  peak (indicated by a solid triangle) and thickness fringes confirming the  growth along the desired pseudo cubic (0 0 1) direction. Out-of-plane pseudocubic lattice    constant ($c_\mathrm{pc}$) for the  SL grown on YAO is found to be 3.875~$\pm$~0.005~\AA, which is  enlarged compared to both bulk ENO and LNO, and is as expected for a tetragonal distortion under high compressive  strain. While the close proximity of the substrate and film peaks for the samples grown on LAO  and  NGO substrates prohibits a reliable estimation of $c_c$, nevertheless it can be immediately  seen  that the film peak for the NGO substrate is close to bulk ENO. Interestingly, $c_\mathrm{pc}$  (3.798 ± 0.005 ~\AA) of the SL grown on DSO is also very close to the lattice constant of bulk ENO (3.8 ~\AA). The very different orbital responses and electronic properties of the SLs (discussed latter in this paper) suggest that such bulk ENO-like lattice constant of the SLs under tensile strain does not arise from simple strain relaxation. We also  note that  single layer films of ENO and LNO under tensile strain also show corresponding bulk-like lattice constants~\cite{jak_prl,eno_prb,icheng_prb} and such anomalous behaviors are related to the strain compensation by  octahedral tilts, rotations and breathing mode distortions.  In contrary to single layer LNO films, LNO layers in the present SLs under tensile strain undergo out-of plane compression  so that the resultant   $c_c$ of the SL remains consistent with the bulk ENO value. This  indicates that the overall symmetry of the SL takes the lower form as in ENO ($a^-a^-c^+$)~\cite{glazernotation}, likely due to the inability of the $a^-a^-a^-$rotation system seen in bulk LNO to stabilize in the presence of the smaller Eu ions.

  \begin{figure}
\includegraphics[width=0.47\textwidth] {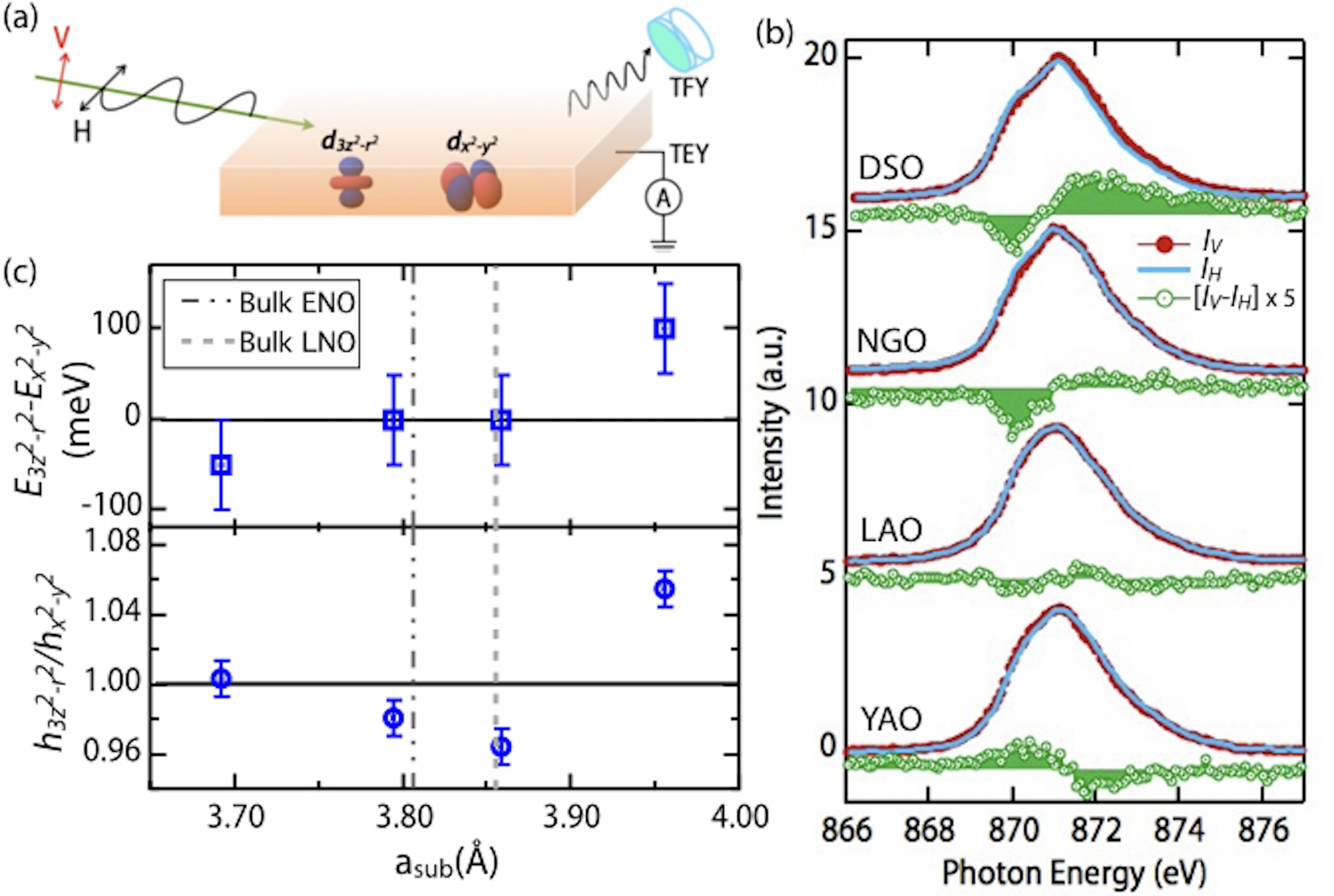}
\caption{\label{} (Color online)  (a)  Experimental geometry of XLD measurement. TEY and TFY refers to total electron yield and total fluorescence yield respectively. (b)    Ni $L_2$ XAS recorded in bulk sensitive TFY mode with horizontally (H) and vertically (V) polarized light and their differences are shown for these 1ENO/1LNO SLs.  Due to strong overlap Ni $L_3$ edge with the La $M_4$ edge,  only Ni $L_2$ edge is  shown. The data have been shifted vertically for clarity. (c) Splitting  and ratio of holes ($X$) between two $e_g$  orbitals have been shown as a function of substrate's in-plane lattice constant.}
\end{figure}

Structural responses of these SLs to the epitaxial strain  have been further investigated by X-ray linear dichroism (XLD) measurements  at room temperature. In such experiment, absorption at Ni $L_{3,2}$ edges are measured with horizontally (H) and vertically (V) polarized X-rays (Fig. 2(a)) and the difference in the energy position and intensity provides information about the  splitting between the $e_g$  orbitals  and their preferential electronic occupation~\cite{jak_prl,lno_epl,keimer_xld13,keimer_xld15,nnoprl,lnoxld}.
 The absorptions labeled as $I_V$ and  $I_H$  after background subtraction of the Ni $L_2$ edge  and the difference signal ($I_V-I_H$) are shown in Fig. 2(b). As seen, the line shapes of the spectra confirm the expected Ni$^{3+}$ oxidation state in these superlattices.  The ratio, $X$ of holes on $d_{3z^2-r^2}$ and $d_{x^2-y^2}$ orbitals can be obtained from the measured $I_V$, $I_H$  using the sum rules
$X=\frac{h_{3z^2-r^2}}{h_{x^2-y^2}}=\frac{3A_V}{[4A_H-A_V]}$~\cite{keimer_xld13,keimer_xld15} where  $A_H$ ($A_V$) is the integrated area of  $I_H$ ($I_V$). We note that bulk $RE$NiO$_3$ does not show any preference between these two $e_g$ levels~\cite{noorbitalordering,co_mazin}.
The variation of  $X$ and the energy splitting between $d_{x^2-y^2}$ and $d_{3z^2-r^2}$  orbitals  obtained from Fig. 2(b)  is plotted  in Fig. 2(c) as a function of substrate lattice constant. As  immediately  seen, the large compressive strain provided by the YAO substrate results in a derivative-like shape of the XLD ($I_V - I_H$) spectra confirming the expected orbital splitting with $c_{pc}$/$a_{pc} >$ 1. $d_{x^2-y^2}$ orbital is higher in energy compared to $d_{3z^2-r^2}$ by 50 meV and $X$ is close to unity, emphasizing equal population on both $e_g$ orbitals.  On the other end,  the SL on DSO with $c_{pc}$/$a_{pc} <$ 1 shows a orbital splitting of around 100 meV and much larger hole density in $d_{3z^2-r^2}$  orbitals.   Surprisingly, for intermediate compressive (LAO) and tensile (NGO) strain cases, the holes density is slightly larger in  $d_{x^2-y^2}$ orbitals and the energy separation between two $e_g$ orbitals is below the accuracy of the XLD measurement ($\sim$ 50 meV).  This apparently conflicting observation for the film on NGO substrate can be resolved by including complex octahedral distortions acting  to accommodate the  moderate amount of strain~\cite{jak_prl,own_enolno}.

  \begin{figure*}
\vspace{-0pt}
\includegraphics[width=1\textwidth] {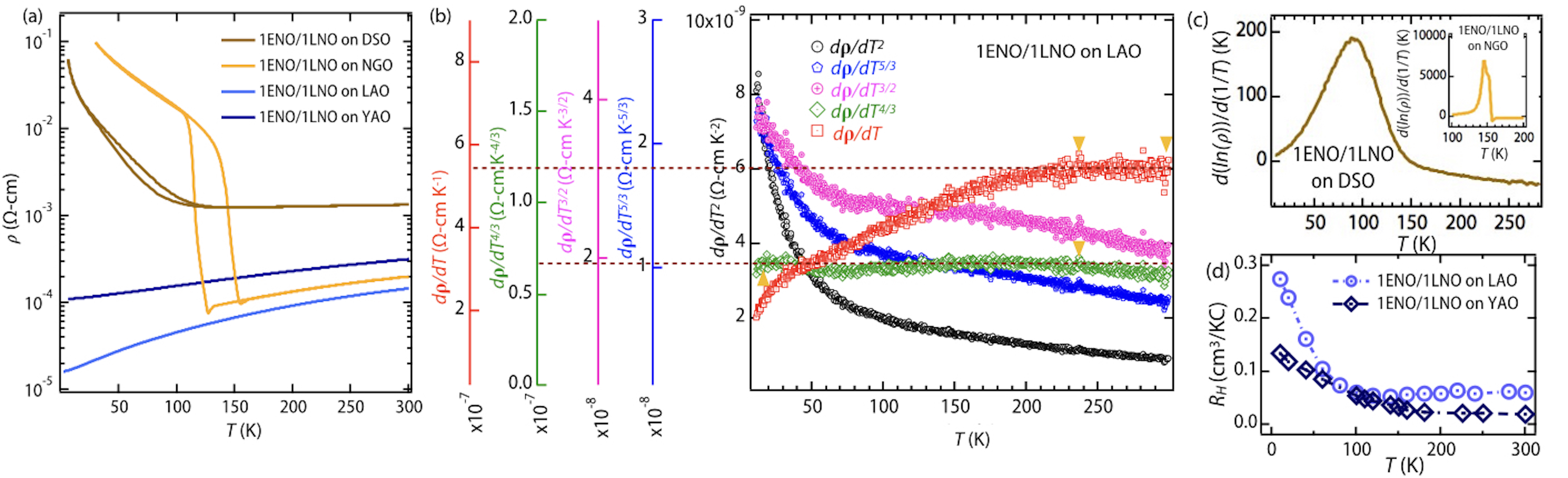}
\caption{\label{} (Color online)  (a) Temperature dependent resistivity of [1ENO/1LNO] superlattice on various substrates. The data of SL on NGO substrate have been adapted from Ref.~\onlinecite{own_enolno}. (b) Resistivity analysis for 1ENO/1LNO SL on LAO substrate.  Yellow triangles indicate the temperature range where the derivative can be considered constant within the noise. As evident from this, $d\rho$/$dT^{4/3}$ is almost temperature-independent (green curve) upto ~230 K. This behaviour changes to linear $T$ dependence after that (red curve).  Similar analysis for other SLs have been shown in Supplemental~\cite{sup}. (c)  Determination of $T_N$ for the film on DSO and NGO substrate by plotting $d(ln\rho)/d(1/T)$ vs. $T$ plot.  (d) Temperature dependence of $R_H$ for SLs grown on LAO and YAO substrate. }
\end{figure*}

The summary of temperature dependent resistivity measurements on these SLs is shown in Fig. 3(a). As reported earlier~\cite{own_enolno}, the samples grown on NGO substrate remains  metallic down to 125 K  and then undergo a MIT.  During heating from low temperature it becomes metallic at 155 K. This hysteresis signifies the first order nature of the transition. This behavior is drastically different from the entirely insulating behavior of single layer ENO or entirely metallic behaviour of LNO films grown on NGO substrates below 300 K~\cite{eno_prb}. Such a large change  emphasizes a complete modulation of the electronic structure by heteroepitaxy. Surprisingly, in  the metallic phase the resistivity  shows an unconventional linear-$T$ dependence (over 190 K - 280 range, shown in Supplemental Materials~\cite{sup}) while the Debye temperature of  bulk nickelates is around 420 K~\cite{debyetemp}.  Such linear-$T$ dependent resistivity has been also observed in high $T_c$ cuprates, pnictide and organic superconductor, ruthenate, heavy fermion metals etc. and has been very often linked to the quantum criticality~\cite{ruthenates,cuprates,pnictides,hightcresistivity,nflexponent}.
 Furthermore, the increase of tensile strain on DSO results  in a higher resistivity at room temperature. Also, while the insulator to metal transition temperature during heating remains similar to the SL on NGO, the magnitude of thermal hysteresis becomes much smaller. On the other hand, the superlattices  grown on LAO and YAO remains metallic down to low temperature without any hysteric behavior.  This suppression of the MIT by compressive strain  resembles the behavior of ultrathin films of PrNiO$_3$, NdNiO$_3$, SmNiO$_3$, EuNiO$_3$~\cite{jian_nc,triscone111,nno_badmetal,eno_prb,sno_aplmaterials,keimer_raman}.  The dc transport of these metallic samples exhibits  a $T^{4/3}$ dependence over a large range of temperature and then switches to linear-$T$ dependent behavior (see Fig. 3(b) and Supplemental Materials~\cite{sup}).    $T^{4/3}$ dependence of resistivity is a characteristic of NFL phase proximal to a two-dimensional quantum critical point~\cite{nflexponent}. Such switching of $T^{4/3}$ dependence to linear $T$ behavior has been also observed in  NdNiO$_3$ thin films under compressive strain and can be accounted by Boltzmann-type transport theory with multiple bands  near a quantum critical point (for details see Refs.~\onlinecite{jian_nc,nflexponent}).

 In the past, long range magnetic orderings of bulk $RE$NiO$_3$, single layer films and superlattice structures consisting of $RE$NiO$_3$ layers have been investigated by neutron diffraction~\cite{etype} and resonant X-ray scattering~\cite{nno_prmaterials,jian_nc,triscone111,keimer_raman,derek_nno,keimer_prl13,noorbitalordering,derek_prb2,scatteringpolycrystal,own_enolno}.  These investigations   showed  that the  insulating phase of these materials always shows a  E$^{\prime}$-antiferromagnetic ordering with the transition temperature  $T_N$ that can be either = $T_\mathrm{MIT}$ or $< T_\mathrm{MIT}$ and the magnetic wave vector is  (1/2, 0, 1/2)$_\mathrm{or}$ [(1/4, 1/4, 1/4)$_\mathrm{pc}$], [$\mathrm{or}$ and $\mathrm{pc}$ denotes orthorhombic and pseudocubic settings respectively]. The signature of this unconventional magnetic ordering can  also be identified in dc transport measurement~\cite{zhou_prl,pno_hall,sno_hall} and  SQUID magnetometry~\cite{rno_squid}. Following the analysis of Zhou et al.~\cite{zhou_prl},  $d(ln\rho)/d(1/T)$ vs. $T$ plot (inset of Fig. 3(c)) was used to determine a  $T_N \sim$  145 K for the SL grown on NGO, which is very close to  $T_N$ (155$\pm$5 K) determined from resonant X-ray scattering measurement~\cite{own_enolno}. Similar analysis for the sample on DSO substrate yields $T_N\sim$ 90 K (Fig. 3(c)). These $T_\mathrm{MIT}$ and $T_N$ are drastically altered compared to bulk compund of formally the same chemical composition Eu$_{0.5}$La$_{0.5}$NiO$_3$ ($T_\mathrm{MIT}$ = $T_N$ = 190K)~\cite{bulk_eudopedlno}. This result implies that  epitaxial strain and the presence of hetero-interface have tremendous impact in ground state in this class of materials.  Further investigations are required to determine any connection between these transition temperatures and orbital polarization of these SLs~\cite{xldtransitiontem}. To examine the possibility for an $E^{\prime}$ antiferromagnetic metallic state~\cite{keimer_raman,lnoafm},  Hall effect measurements were carried out on the metallic superlattices. Previous work on     nickelates  indicates that  Hall coefficient ($R_H$) shows a sign change from hole like to electron like behavior  around $T_N$ ~\cite{pno_hall,sno_hall}. This sign switching behaviour is absent in our SL s (see Fig. 3(d)) and this indicates that these  metallic SLs do not  likely to have E$^{\prime}$-AFM ordering.


In general, electronic structure of correlated materials is parametrized by hopping strength ($t$), electron-electron correlation ($U$), and charge transfer energy ($\Delta$) or effective charge transfer energy ($\Delta^\prime$) in context of the  Zannen-Sawatzky-Allen (ZSA) phase diagram~\cite{zsaphase} or its' modified version~\cite{cuo2_dd}. In relation to nickelates, very early photoemission spectroscopy measurements revealed that $RE$NiO$_3$ has very small charge transfer energy~\cite{pnoxas,xps_dd}. The insulating phase has been  identified as a covalent insulator by Barman et al~\cite{xps_dd} with the gap arising from the $d^8\underline{L}$+ $d^8\underline{L}\rightarrow d^8$+ $d^8\underline{L}^2$ charge fluctuations~\cite{cuo2_dd,nco_fujimori} (here $\underline{L}$ denotes a hole on oxygen $p$ orbital). The importance of  ligand hole states in realizing the insulator to metal transition in $RE$NiO$_3$  has been further emphasized in several recent theoretical and experimental works~\cite{khomskii_holeordering,millis_siteselective,sawatzky_hf,georges_dmft,own_enolno}. In addition, a recent RIXS (resonant inelastic x-ray scattering) experiment on Ni has clearly confirmed the presence of negative $\Delta^\prime$ and the band gap of a O 2$p$-2$p$ type~\cite{rixs}. To understand the strain induced suppression of the insulating phase, we focus on O $K$-edge resonant X-ray absorption  spectra~\cite{xas1,pnoxas,jian_nc,nidopedlio,cupratexas,john}, where ligand hole states ($d^8\underline{L}$) can be identified   as a prepeak  around 528.5 eV due to the $d^8\underline{L}\rightarrow\underline{c}d^8$ transition (here $\underline{c}$ is a hole in oxygen 1$s$ core state) The degree of Ni-O bond covalency can be monitored by the intensity, position, and width of this prepeak.

 \begin{figure}
\vspace{-0pt}
\includegraphics[width=.48\textwidth] {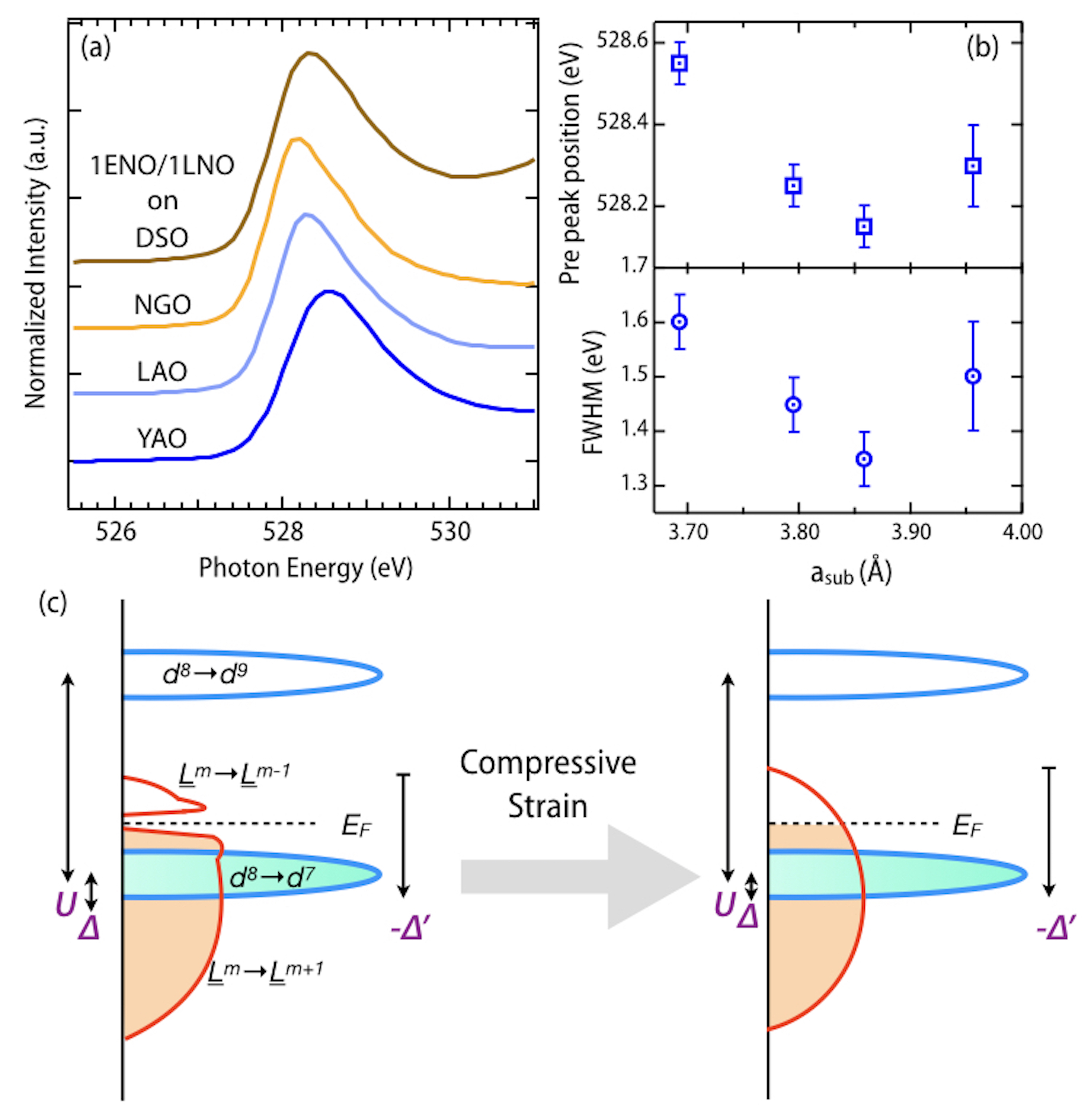}
\caption{\label{} (Color online)   (a) Pre peak of O $K$-edge  absorption around 528.5 eV for 1ENO/1LNO SLs measured at 300 K. The peaks above 530 eV have been shown in supplemental information. (b) Energy shift and FWHM of the pre-peak as a function of  substrate's in-plane lattice constant. An additional strong peak, related to the transition to hybridized Sc 3$d$-O 2$p$ states is present around 533 eV for 1ENO/1lNO SL on DSO (see Supplemental Information~\cite{sup}).  (c) Schematic representation of the single-particle density of states  in terms of charge removal and charge addition  for a negative charge transfer material with $d^8\underline{L}^m$ as ground state, adapted from Ref.~\onlinecite{rixs}. Left and right panel corresponds to covalent insulator and  $pd$ metal.}
\end{figure}

  A direct inspection of Figure 4(a) and upper panel of Fig. 4(b) shows the movement of the prepeak towards higher photon energy as  strain becomes more compressive (NGO $\rightarrow$ LAO $\rightarrow $YAO), thus emphasizing a decrease of charge transfer energy $\Delta$ with compressive strain.  Microscopically,  at the first  approximation $\Delta$ is related to the electron affinity of oxygen [$I$(O$^{2-}$)], the   ionization potential of  Ni$^{3+}$ [$A$(Ni$^{3+}$)],  relative Madelung potential $\delta V_\mathrm{Mad}$ between Ni and O, and the nearest-neighbor distance between Ni and O ($d_\mathrm{Ni-O}$) as $\Delta$ = $e \delta V_\mathrm{Mad}$ + $I$(O$^{2-}$) - $A$(Ni$^{3+}$) - $e^2$/$d_\mathrm{Ni-O}$~\cite{mit_rmp}.  This  observation implies  that strain induced change in $\Delta$ originates from the strong modulation in the relative Madelung potential. Most importantly, the FWHM of the pre-peak also increases with compressive strain signifying the enhancement of Ni-O hybridization. The modulation in both charge transfer energy and hybridization (covalency) results in the complete suppression of the insulating phase as schematically illustrated  in Fig. 4(c). Such strain-induced `self-doping' of 1ENO/1LNO SL highlights the  utility of strain as a means of effective carrier doping without detrimental effects of chemical disorders.

In contrast to  the SLs  on other substrates, 1ENO/1LNO SL on DSO has a sizable energy splitting between two $e_g$ orbitals with different electronic occupancies (Fig. 2(c)). This implies the dominance of a uniform Jahn-Teller distortion over the other structural distortion modes for this SL (e.g. breathing mode, staggered Jahn-Teller order etc~\cite{landautheorymillis}).  In this  case, the electronic transitions from O 1$s\rightarrow$ Ni $d_{3z^2-r^2}$-O $p_z$ hybridized states and O 1$s\rightarrow$ Ni $d_{x^2-y^2}$-O $p_x$, $p_y$ hybridized states occur at slightly different energies and with different intensity (intensity $\propto$ number of holes). This in turn can lead to  the observed shift of O $K$ pre-peak to higher photon energy and corresponding enhancement in  FWHM for the SL on DSO (Fig. 3c). Further RIXS and, polarized XAS experiment on O $K$ edge will be required to investigate this scenario~\cite{rixsdm}.

\section{Conclusion}
To summarize, correlated metal LaNiO$_3$ and charge transfer insulator EuNiO$_3$ have been heterostructured in the form of unit cell superlattices 1 uc EuNiO$_3$/1 uc LaNiO$_3$  and  the effects of epitaxial strain have been investigated using XRD, dc transport, Hall effect, resonant XAS and XLD measurements. The electronic and magnetic phases are highly tunable by  application of strain and several unusual  phases including non-Fermi liquid, paramagnetic insulator, antiferromagnetic insulator phases have been observed. The detailed analysis of XAS spectra on oxygen $K$-edge revealed  strain induced strong modulation of charge transfer energy and covalency, resulting  insulator to metal transition.

\section{Acknowledgements}
 S.M. and J.C. deeply thank D. Khomskii and  K. Haule for theoretical discussion. S.M. is supported by IISc start up grant.  D.M. is  supported by the U.S. Department of Energy, Office of Basic Energy Sciences, Early Career Award Program under Award No.  1047478.  Work at Brookhaven National Laboratory was supported by the U.S. Department of Energy,  Office of Science, Office of Basic Energy Sciences, under Contract No. DESC0012704.  J.C., X.L.   are supported by the Gordon and Betty Moore Foundation EPiQS Initiative through Grant No. GBMF4534.  X.L.  acknowledges support of DE-SC 00012375 grant for synchrotron  work. This research used resources of the Advanced Photon Source, a U.S. Department of Energy Office of Science User Facility operated by Argonne National Laboratory under Contract No. DE-AC02-06CH11357.


\begin{thebibliography}{99}
\bibitem{mit_rmp} M. Imada, A. Fujimori, and Y. Tokura, Rev.  Mod. Phys. \textbf{70}, 1039 (1998).
\bibitem{ramanathanreview} Z. Yang, C. Ko, and S. Ramanathan, Annu. Rev. Mater. Res.  {\bf 41}, 337 (2011).
\bibitem{tokuranature} M. Nakano, K. Shibuya, D. Okuyama, T. Hatano, S. Ono, M. Kawasaki, Y. Iwasa, and Y. Tokura, Nature, {\bf 487}, 459 (2012).
\bibitem{parkinscience}	J. Jeong, N. Aetukuri, T. Graf, T. D. Schladt, M. G. Samant, and S. S. P. Parkin Science  {\bf 339}, 1402 (2013). 
\bibitem{schlom} D. G. Schlom, L.-Q. Chen, X. Pan, A. Schmehl, and M. A. Zurbuchen, J. Am. Ceram. Soc.  {\bf 91}, 2429 (2008).
\bibitem{hwang} H. Y. Hwang, Y. Iwasa, M. Kawasaki, B. Keimer, N. Nagaosa, and Y. Tokura, Nat. Mater.  {\bf 11}, 103 (2012).
\bibitem{jakrmp} J. Chakhalian, J.W. Freeland, A. J. Millis, C. Panagopoulos, and J. M. Rondinelli, Rev. Mod. Phys.  {\bf 86}, 1189 (2014).

\bibitem{ownreview} S. Middey, J. Chakhalian, P. Mahadevan, J. W. Freeland, A. J. Millis, and D. D. Sarma, Annu. Rev. Mater. Res. {\bf 46}, 305-334 (2016).
\bibitem{ramanathan1} Y. Zhou, X. Guan, H. Zhou, K. Ramadoss, S. Adam, H. Liu, S. Lee, J. Shi, M. Tsuchiya, D. D. Fong and S. Ramanathan, Nature, 2016, 231, 534.
\bibitem{ramanathan2} Z. Zhang, D. Schwanz, B. Narayanan, M. Kotiuga, J. A. Dura, M. Cherukara, H. Zhou, J. W. Freeland, J. Li, R. Sutarto, F. He, C. Wu, J. Zhu, Y. Sun, K. Ramadoss, S. S. Nonnenmann, N. Yu, R. Comin, K. M. Rabe, S. K. R. S. Sankaranarayanan, and  S. Ramanathan, Nature, 2018, 553, 68.


\bibitem{jian_nc} J. Liu,  M. Kargarian, M. Kareev, B. Gray, P. J. Ryan, A. Cruz, N. Tahir, Yi-De Chuang, J. Guo, J. M. Rondinelli, J. W. Freeland, G. A. Fiete, and  J. Chakhalian, Nat. Commun. {\bf 4}, 2714  (2013).
\bibitem{triscone111} S. Catalano, M. Gibert, V. Bisogni, F. He, R. Sutarto, M. Viret, P. Zubko, R. Scherwitzl, G. A. Sawatzky, T. Schmitt, and J.-M. Triscone, APL Materials {\bf 3}, 062506 (2015).
\bibitem{nno_badmetal}E. Mikheev, A. J. Hauser, B. Himmetoglu, N. E. Moreno, A. Janotti, C. G. Van de Walle, and S. Stemmer, Sci. Adv. {\bf 1}, e1500797 (2015).
 \bibitem{eno_prb} D.  Meyers,  S. Middey, M. Kareev, M. van Veenendaal, E. J. Moon, B. A. Gray, Jian Liu, J. W. Freeland, and J. Chakhalian, Phys. Rev. B {\bf 88}, 075116 (2013).
 \bibitem{sno_aplmaterials} S. Catalano, M. Gibert, V. Bisogni, O. E. Peil, F. He,   R. Sutarto, M. Viret, P. Zubko, R. Scherwitzl,   A. Georges, G. A. Sawatzky, T. Schmitt, and J.-M. Triscone, APL Materials {\bf 2}, 116110 (2014).
\bibitem{keimer_raman}M. Hepting,  M. Minola,  A. Frano,  G. Cristiani,  G. Logvenov,  E. Schierle,  M. Wu,  M. Bluschke, E. Weschke,  H.-U. Habermeier,  E. Benckiser,  M. Le Tacon,  and B. Keimer, Phys. Rev. Lett. {\bf 113}, 227206 (2014).
 \bibitem{sno_prb} F. Y. Bruno, K. Z. Rushchanskii, S. Valencia, Y. Dumont, C. Carr$\acute{e}$t$\acute{e}$ro, E. Jacquet, R. Abrudan, S. Bl$\ddot{u}$gel, M. Le$	\breve{z}$ai$\acute{C}$, M. Bibes, and A. Barth$\acute{e}$l$\acute{e}$my, Phys. Rev. B {\bf 88}, 195108 (2013).


\bibitem{snojap} G. H. Aydogdua, S. D. Ha, B. Viswanath, and S. Ramanathan, Journal of Applied Physics {\bf 109}, 124110 (2011).


\bibitem{lnothoery} D. Puggioni, A. Filippetti, and V. Fiorentini, Phys. Rev. B {\bf 86}, 195132 (2012).
 
\bibitem{lnoarpes} Hyang Keun Yoo, Seung Ill Hyun, Young Jun Chang, Luca Moreschini, Chang Hee Sohn, Hyeong-Do Kim, Aaron Bostwick, Eli Rotenberg, Ji Hoon Shim, and Tae Won Noh
Phys. Rev. B 93, 035141 (2016).



 \bibitem{nno_co} U. Staub, G. I. Meijer, F. Fauth, R. Allenspach, J. G. Bednorz, J. Karpinski, S. M. Kazakov, L. Paolasini, and F. d'Acapito, Phys. Rev. Lett. {\bf 88}, 126402 (2002).
\bibitem{scagnoliprb} V. Scagnoli, U. Staub, M. Janousch, A. M. Mulders, M. Shi, G. I. Meijer, S. Rosenkranz, S. B. Wilkins, L. Paolasini, J. Karpinski, S. M. Kazakov, and S. W. Lovesey,  Phys. Rev. B {\bf 72}, 155111 (2005).
\bibitem{lorenzoprb} J. E. Lorenzo, J. L. Hodeau, L. Paolasini, S. Lefloch, J. A. Alonso, and G. Demazeau,  Phys. Rev. B {\bf 71}, 045128 (2005).
\bibitem{noorbitalordering} V. Scagnoli, U. Staub, A. M. Mulders, M. Janousch, G. I. Meijer, G. Hammerl, J. M. Tonnerre, and N. Stojic, Phys. Rev. B {\bf 73}, 100409(R) (2006).
\bibitem{keimer_scattering} Y. Lu, A. Frano, M. Bluschke, M. Hepting, S. Macke, J. Strempfer, P. Wochner, G. Cristiani, G. Logvenov, H.-U. Habermeier, M. W. Haverkort, B. Keimer, and E. Benckiser, Phys. Rev. B {\bf 93}, 165121 (2016).




\bibitem{nno_upton} M. Upton, Y. Choi, H. Park, Jian Liu, D. Meyers, J. Chakhalian, S. Middey, J.-W. Kim, P. J. Ryan, Phys. Rev. Lett. {\bf 115}, 036401 (2015).
\bibitem{derek_nno} D. Meyers, J. Liu, J.W. Freeland, S. Middey, M. Kareev, J. Kwon, J.M. Zuo, Y.-D. Chuang, J. W. Kim, P.J. Ryan, and J. Chakhalian, Sci. Rep. {\bf 6}, 27934 (2016).

\bibitem{lno_keimer_nm}E. Benckiser, M. W. Haverkort, S. Br\"{u}ck, E. Goering, S. Macke, A. Frano, X. Yang, O. K. Andersen, G. Cristiani, H. -U. Habermeier, A. V. Boris, T. Zegkinoglou, P. Wochner, H.-J. Kim, V. Hinkov and  B. Keimer, Nat. Mater. {\bf 10},  189 (2011).
\bibitem{lno_jian_prb}J. Liu, S. Okamoto, M. van Veenendaal, M. Kareev, B. Gray, P. Ryan, J. W. Freeland, and J. Chakhalian, Phys. Rev. B {\bf 83}, 161102(R) (2011).
\bibitem{lno_keimer_science}A. V. Boris, Y. Matiks, E. Benckiser, A. Frano, P. Popovich, V. Hinkov, P. Wochner, M. Castro-Colin, E. Detemple,  V. K. Malik, C. Bernhard, T. Prokscha, A. Suter, Z. Salman, E. Morenzoni, G. Cristiani, H.-U. Habermeier, B. Keimer, Science {\bf 332}, 937 (2011).
\bibitem{keimer_prl13}  A. Frano, E. Schierle, M. W. Haverkort, Y. Lu, M. Wu, S. Blanco-Canosa, U. Nwankwo, A. V. Boris, P. Wochner, G. Cristiani, H. U. Habermeier, G. Logvenov, V. Hinkov, E. Benckiser, E. Weschke, and B. Keimer,   Phys. Rev. Lett.  {\bf 111}, 106804 (2013).
\bibitem{keimer_xld13}M. Wu, E. Benckiser, M. W. Haverkort, A. Frano, Y. Lu, U. Nwankwo, S. Brï¿½ck, P. Audehm, E. Goering, S. Macke, V. Hinkov, P. Wochner, G. Christiani, S. Heinze, G. Logvenov, H.-U. Habermeier, and B. Keimer Phys. Rev. B {\bf 88}, 125124 (2013).
\bibitem{lno_epl} J. W. Freeland,  J. Liu,  M. Kareev, B. Gray,  J. W. Kim, P. Ryan, R. Pentcheva, and J. Chakhalian, Euro Phys. Lett. {\bf 96}, 57004 (2011).
\bibitem{keimer_xld15}M. Wu, E. Benckiser, P. Audehm, E. Goering, P. Wochner, G. Christiani, G. Logvenov, H.-U. Habermeier, and B. Keimer, Phys. Rev. B {\bf 91}, 195130 (2015).
\bibitem{nno_prmaterials}A. S. Disa, A. B. Georgescu, J. L. Hart, D. P. Kumah, P. Shafer, E. Arenholz, D. A. Arena, S. Ismail-Beigi, M. L. Taheri, F. J. Walker, and Charles H. Ahn, Phys. Rev. Materials, {\bf 1}, 024410 (2017).

\bibitem{own_enolno} S. Middey, D. Meyers, M. Kareev, Y. Cao, X. Liu, P. Shafer, J. W. Freeland, J. W. Kim, P. J. Ryan, and J. Chakhalian, Phys. Rev. Lett. {\bf 120}, 156801 (2018).

\bibitem{rno_phase1}M. L. Medarde, J. Phys.: Condens. Matter {\bf 9}, 1679 (1997).
\bibitem{rno_phase} G. Catalan,   {\it Phase Transit.} {\bf 81}, 729 -749 (2008).

\bibitem{bulk_eudopedlno} R. D. S$\acute{a}$nchez, M. T. Causa, A. Seoane, J. Rivas, F. Rivadulla, and M. A. L$\acute{o}$pez-Quintela, J. J. P$\acute{e}$rez Cacho, J. Blasco, and J. Garc\'{i}a, Journal of Solid State Chemistry {\bf 151}, 1 (2000).


\bibitem{icheng_prb}I. C. Tung, P. V. Balachandran, Jian Liu, B. A. Gray, E. A. Karapetrova, J. H. Lee, J. Chakhalian, M. J. Bedzyk, J. M. Rondinelli, and J. W. Freeland Phys. Rev. B {\bf 88}, 205112 (2013).
\bibitem{selfdoping} M. A. Korotin, V. I. Anisimov, D. I. Khomskii, and G. A. Sawatzky, Phys. Rev. Lett. {\bf 80}, 4305 (1998).
\bibitem{selfdopinglno} E. J. Moon, J. M. Rondinelli, N. Prasai, B. A. Gray, M. Kareev, J. Chakhalian, and J. L. Cohn, Phys. Rev. B {\bf 85}, 121106(R) (2012).
  \bibitem{misha_jap}M. Kareev, S. Prosandeev, B. Gray, J. Liu, P. Ryan, A. Kareev, E. J. Moon, and J. Chakhalian,  J. Appl. Phys. {\bf 109}, 114303 (2011).
 \bibitem{nnoprl} S. Middey, D. Meyers, D. Doennig, M .Kareev, X. Liu, Y. Cao, Z. Yang, J. Shi, L. Gu, P. J. Ryan, R. Pentcheva, J. W. Freeland, and J. Chakhalian, Phys. Rev. Lett. 1{\bf 16}, 056801 (2016).

\bibitem{scirep} S. Middey,  P. Rivero, D. Meyers,  M. Kareev, X. Liu, Y. Cao, J. W. Freeland, S. Barraza-Lopez and J. Chakhalian, Sci. Rep. {\bf 4}, 6819 (2014).
\bibitem{eno_growth}D. Meyers, E. J. Moon, M. Kareev, I. C. Tung, B. A. Gray, J. Liu, M. J. Bedzyk, J.W. Freeland, and J. Chakhalian, J. Phys. D: Appl. Phys. (2013).
\bibitem{own_apl} S. Middey,  D. Meyers,  M. Kareev,  E. J. Moon,  B. A. Gray, X. Liu,  J. W. Freeland, and J. Chakhalian, Appl. Phys. Lett. {\bf 101}, 261602 (2012).
\bibitem{jak_prl} J. Chakhalian, J. M. Rondinelli, Jian Liu, B. A. Gray, M. Kareev, E. J. Moon, N. Prasai, J. L. Cohn, M. Varela, I. C. Tung, M. J. Bedzyk, S. G. Altendorf, F. Strigari, B. Dabrowski, L. H. Tjeng, P. J. Ryan, and J. W. Freeland, Phys. Rev. Lett. {\bf 107}, 116805 (2011).
\bibitem{glazernotation} A. M. Glazer, Acta Cryst. {\bf B28}, 3384  (1972).

\bibitem{lnoxld} A. S. Disa, F. J. Walker, S. Ismail-Beigi, and C. H. Ahn, APL Materials {\bf 3}, 062303 (2015).
\bibitem{co_mazin}I. I. Mazin, D. I. Khomskii, R. Lengsdorf, J. A. Alonso, W. G. Marshall, R. M. Ibberson, A. Podlesnyak, M. J. Mart\'{i}nez-Lope, and M. M. Abd-Elmeguid, Phys. Rev. Lett. {\bf 98}, 176406 (2007).
\bibitem{sup} See Supplemental Material.
\bibitem{debyetemp} K. P.  Rajeev, G. V. Shivashankar and A. K. Raychaudhuri,  Solid State Communications, {\bf 79},  591,  (1991).
\bibitem{ruthenates}J. A. N. Bruin, H. Sakai, R. S. Perry, and A. P. Mackenzie, Science, {\bf 339}, 804 (2013).
\bibitem{cuprates}R. A. Cooper, Y. Wang, B. Vignolle, O. J. Lipscombe, S. M. Hayden, Y. Tanabe, T. Adachi,  Y. Koike,  M. Nohara, H. Takagi,  Cyril Proust, and N. E. Hussey, Science {\bf 323}, 603 (2009).
\bibitem{pnictides}S. Kasahara,  T. Shibauchi,  K. Hashimoto,  K. Ikada,  S. Tonegawa,  R. Okazaki,  H. Shishido,  H. Ikeda,  H. Takeya, K. Hirata,  T. Terashima,  and Y. Matsuda, Phys. Rev. B {\bf 81}, 184519 (2010).
\bibitem{hightcresistivity} T. M. Rice, N. J. Robinson,  and Al. M. Tsvelik, Phys. Rev. B {\bf  96}, 220502(R) (2017).
\bibitem{nflexponent}D.  L. Maslov,  V.  I.  Yudson, and A. V. Chubukov,  Phys. Rev. Lett. {\bf 106}, 106403 (2011).



\bibitem{etype} J. L. Garc\'{i}a-Mu\~{n}oz, , J. Rodr\'{i}guez-Carvajal, and P. Lacorre, Phys. Rev. B {\bf 50}, 978 (1994).
\bibitem{derek_prb2} D. Meyers, S. Middey, M. Kareev, Jian Liu, J. W. Kim, P. Shafer, P. J. Ryan, and J. Chakhalian, Phys. Rev. B {\bf 92}, 235126 (2015).
\bibitem{scatteringpolycrystal}Y. Bodenthin, U. Staub, C. Piamonteze, M. Garcï¿½a-Fernï¿½ndez, M. J. Martï¿½nez-Lope, and J. A. Alonso, J. Phys. Condens. Matter 23, 036002 (2011).

 \bibitem{zhou_prl}J.-S. Zhou, J. B. Goodenough, and B. Dabrowski, Phys. Rev. Lett. {\bf 95}, 127204 (2005).
\bibitem{pno_hall} S. W. Cheong, H. Y. Hwang, B. Batlogg, A. S. Cooper, and P. C. Canfield, Physica B {\bf 194-196}, 1087 (1994).
 \bibitem{sno_hall}S. D. Ha, R. Jaramillo,  D. M. Silevitch,  F. Schoofs,  K. Kerman,  J. D. Baniecki,  and S. Ramanathan, Phys. Rev. B {\bf 87}, 125150 (2013).
 \bibitem{rno_squid} J.-S. Zhou, J. B. Goodenough, B. Dabrowski, P.W. Klamut, and Z. Bukowski, Phys. Rev. Lett., {\bf 84}, 526  (2000).

\bibitem{xldtransitiontem}J. J. Peng, C. Song, M. Wang, F. Li, B. Cui, G. Y. Wang, P. Yu, and F. Pan, Phys. Rev. B {\bf 93}, 235102 (2016).
\bibitem{lnoafm}H. Guo, Z. W. Li, L. Zhao, Z. Hu, C. F. Chang, C. -Y.  Kuo, W. Schmidt, A. Piovano, T.  W. Pi, O. Sobolev, D. I.  Khomskii, L. H. Tjeng and A. C. Komarek, Nature Communications {\bf 9}, 43 (2018).




\bibitem{zsaphase} J. Zaanen, G. A. Sawatzky, J. W. Allen, Phys. Rev. Lett. {\bf 55}, 418 (1985).
\bibitem{cuo2_dd}S. Nimkar, D. D. Sarma,  H. R. Krishnamurthy,  and S. Ramasesha,  Phys. Rev. B {\bf 48}, 7355-7363 (1993). 
\bibitem{xps_dd} S. R. Barman, A. Chainani, and D. D. Sarma, Phys. Rev. B {\bf 49}, 8475 (1994).
\bibitem{pnoxas}T. Mizokawa, A. Fujimori, T. Arima, Y. Tokura, N. Mori, and J. Akimitsu, Phys. Rev. B {\bf 52}, 13865 (1995).
\bibitem{nco_fujimori}T. Mizokawa, H. Namatame, and A. Fujimori, K. Akeyama, H. Kondoh, and H. Kuroda, and N. Kosugi, Phys. Rev. B {\bf 67}, 1638 (1991).


\bibitem{khomskii_holeordering} T. Mizokawa, D. I. Khomskii, and G. A. Sawatzky, Phys. Rev. B {\bf 61}, 11263 (2000).
\bibitem{millis_siteselective}H. Park,  A. J. Millis,  and C. A. Marianetti, Phys. Rev. Lett. {\bf 109}, 156402 (2012).
\bibitem{sawatzky_hf}S. Johnston, A. Mukherjee,  Ilya Elfimov, M. Berciu,  and G. A. Sawatzky, Phys. Rev. Lett. {\bf 112}, 106404 (2014).
\bibitem{georges_dmft}A. Subedi,  Ol. E. Peil, and A. Georges, Phys. Rev. B {\bf 91}, 075128 (2015).
\bibitem{rixs} V. Bisogni, S. Catalano, R. J. Green, M. Gibert, R. Scherwitzl, Y. Huang, V. N. Strocov, P. Zubko, S. Balandeh, J.-M. Triscone, G. Sawatzky, and T. Schmitt, Nat. Commun. {\bf 7}, 13017 (2016).
\bibitem{xas1}J. Garc\'{i}a, J. Blasco, M. G. Proietti, and M. Benfatto, Phys. Rev. B {\bf 52}, 15823 (1995).
\bibitem{nidopedlio} P. Kuiper, G. Kruizinga, J. Ghijsen, G. A. Sawatzky, and H. Verweij, Phys. Rev. Lett. {\bf 62}, 221(1989).
\bibitem{cupratexas}N. Nucker, H. Romberg, X. X. Xi, J. Fink, B. Gegenheimer, and Z. X. Zhao, Phys. Rev. B, {\bf 39}, 6619 (1989).
\bibitem{john}J. W.Freeland, M. van Veenendaal, and J. Chakhalian, Journal of Electron Spectroscopy and Related Phenomena, {\bf 208},  56 (2016).

\bibitem{landautheorymillis} Z. He, and A. J. Millis, Phys. Rev. B B {\bf 91}, 195138 (2015).
\bibitem{rixsdm} G. Fabbris, D. Meyers, J. Okamoto, J. Pelliciari, A. S. Disa, Y. Huang, Z.-Y. Chen, W.?B. Wu, C. T. Chen, S. Ismail-Beigi, C. H. Ahn, F. J. Walker, D. J. Huang, T. Schmitt, and M. P. M. Dean, Phys. Rev. Lett. {\bf 117}, 147401 (2016).


 \end{thebibliography}
 \end{document}